\documentclass[10pt,conference]{IEEEtran}
\usepackage{array}
\usepackage{booktabs}
\usepackage{graphicx}
\usepackage{subfigure}
\usepackage{wrapfig}
\usepackage[ruled]{algorithm2e}
\usepackage{multirow}
\usepackage{threeparttable}
\usepackage{amsmath}
\usepackage{amssymb}

\begin{document}

\title{Lightweight and Generalizable Multi-Sensor Human Activity Recognition via Cascaded Fusion and Style-Augmented Decomposition}

\author{
Wang Chenglong$^{1,\ast}$,
Zhuo Yan$^{1,\ast}$,
Ding Wenbo$^{1}$,
and Chen Xinlei$^{1,\dagger}$~\IEEEmembership{Member,~IEEE}

        % <-this % stops a space

%\thanks{This work was supported in part by the National Natural Science Foundation of China (Grant no. 62101382), in part by Key Research \& Development and Transformation Project of Qinghai Province (Grant no. 2022-QY-212), and in part by the Fundamental Research Funds for the Central Universities (Grant no. 2242022R10023).}

% \thanks{Y. Zhuo and X. L Chen are with the Shenzhen International Graduate School, Tsinghua University, Shenzhen, China (e-mail: zy23@mails.tsinghua.edu.cn; chen.xinlei@sz.tsinghua.edu.cn).}

% \thanks{J. F Zheng is with the Shenzhen Smartcity Communication, Shenzhen, China (e-mail: zhengjianfeng@smartcitysz.com).}

% \thanks{X. Z Kong is with the Beijing University of Aeronautics and Astronautics, Beijing, China (e-mail: 22373248@buaa.edu.cn).}

%\thanks{This paper was produced by the IEEE Publication Technology Group. They are in Piscataway, NJ.}% <-this % stops a space
%\thanks{Manuscript received April 19, 2021; revised August 16, 2021.}
}

\markboth{Journal of \LaTeX\ Class Files,~Vol.~14, No.~8, August~2021}%
{Shell \MakeLowercase{\textit{et al.}}: A Sample Article Using IEEEtran.cls for IEEE Journals}


\maketitle

\begin{abstract}
Wearable Human Activity Recognition (WHAR) is a prominent research area within ubiquitous computing, whose core lies in effectively modeling intra- and inter-sensor spatio-temporal relationships from multi-modal time series data. Existing methods either suffer from high computational complexity due to attention-based fusion or lack robustness to data variations during feature extraction. To address these issues, we propose a lightweight and generalizable framework that retains the core "decomposition-extraction-fusion" paradigm while introducing two key innovations. First, we replace the computationally expensive Attention and Cross-Variable Fusion (CVF) modules with a Cascaded Fusion Block (CFB), which achieves efficient feature interaction without explicit attention weights through the operational process of "compression-recursion-concatenation-fusion". Second, we integrate a MixStyle-based data augmentation module before the Local Temporal Feature Extraction (LTFE) and Global Temporal Aggregation (GTA) stages. By mixing the mean and variance of different samples within a batch and introducing random coefficients to perturb the data distribution, the model's generalization ability is enhanced without altering the core information of the data. The proposed framework maintains sensor-level, variable-level, and channel-level independence during the decomposition phase, and achieves efficient feature fusion and robust feature extraction in subsequent processes. Experiments on two benchmark datasets (Realdisp, Skoda) demonstrate that our model outperforms state-of-the-art methods in both accuracy and macro-F1 score, while reducing computational overhead by more than 30\% compared to attention-based baselines. This work provides a practical solution for WHAR applications on resource-constrained wearable devices.
\end{abstract}


\begin{figure*}
\centering
\includegraphics[scale = 0.4]{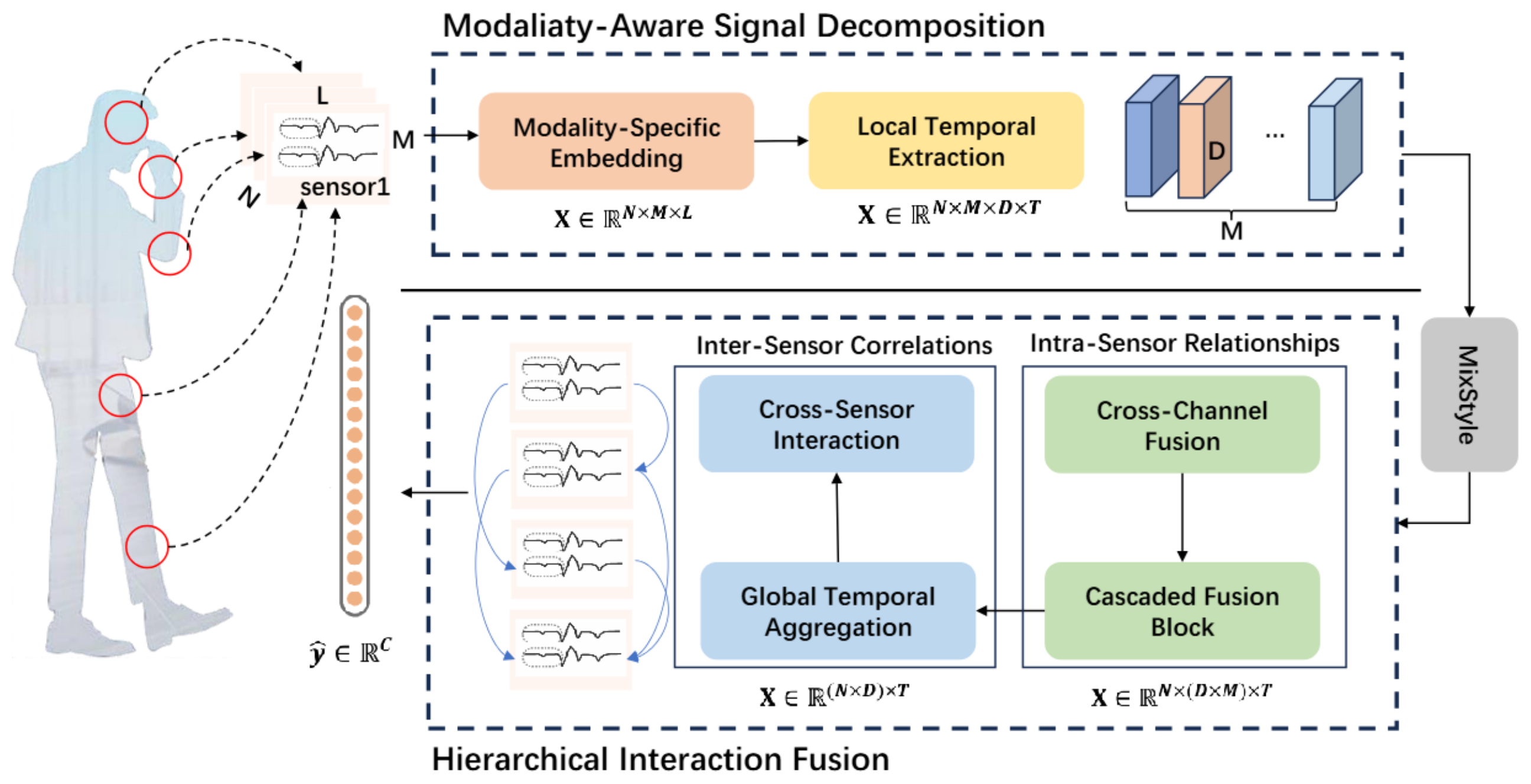}
\caption{Architecture of Our Model.}
\label{fig:title}
\end{figure*}

% \begin{figure*}
% \centering
% \includegraphics[scale = 0.65,clip=true, trim=0 210 60 40]{Fig/datatochip.pdf}
% \caption{The overall pipeline of the proposed method.}
% \label{fig:overallpipeline}
% \end{figure*}

\section{Introduction}
Wearable Human Activity Recognition (WHAR) has become an indispensable part of pervasive computing, with far-reaching applications across multiple domains. It plays a crucial role in medical care, security, entertainment, and tactical scenarios\cite{lara2012survey}, and also finds extensive use in elderly care, clinical monitoring, and smart environments\cite{uddin2021human,kanjilal2025human}. Moreover, it has made significant inroads into mass sports activities, promoting the development of activity evaluation systems\cite{liang2023application}. The significance of WHAR lies in its ability to provide accurate and timely information on human activities and behaviors, serving as the foundation for intelligent supportive systems in various fields\cite{lara2012survey}. For example, in smart eldercare environments, WHAR systems based on wearable sensors can monitor the elderly, warn of fall risks, and extend their independent living time\cite{uddin2021human}. In clinical settings, real-time gait analysis through WHAR contributes to advanced diagnosis and treatment support\cite{kalpana2025explainable}. With the aging population trend, WHAR has emerged as a promising assistive technology to support the daily lives of older individuals\cite{ni2024survey}.

The advancement of WHAR is driven by three core pillars: acquisition devices, artificial intelligence (AI) techniques, and diverse applications\cite{liu2021overview}. Acquisition devices include smartphones, wearable IoT devices, and inertial measurement units (IMUs)\cite{bianchi2019iot}, which integrate multiple sensors such as accelerometers, gyroscopes, and barometers to capture comprehensive activity data\cite{alarfaj2024wearable}. AI techniques, especially deep learning, have revolutionized WHAR by enabling automatic feature extraction and accurate classification\cite{zhang2022deep}. However, multi-sensor data modeling in WHAR faces several core demands and challenges. Firstly, it needs to handle noisy and diverse data to ensure consistency and accuracy in data collection across varying activities and environments\cite{kanjilal2025human}. Secondly, real-time performance is crucial, requiring reduced computational overhead while maintaining high accuracy\cite{kalpana2025explainable}. Thirdly, addressing data heterogeneity and privacy concerns is essential when dealing with distributed user data\cite{chai2024profile}. Fourthly, capturing spatial and temporal dependencies in time-series sensor data is vital for precise activity classification\cite{sheker2024wearable}. Finally, enhancing model interpretability and reducing reliance on large labeled datasets are also key requirements\cite{ghorrati2025scalable}.

To meet these demands, numerous studies have focused on multi-sensor data fusion techniques. UC Fusion was proposed to integrate unique sensor features with common features of all sensors\cite{jiang2015human}. It unifies data formats through segmentation and dimensional transformation, effectively handling heterogeneous data\cite{jiang2015human}. The late-fusion technique has also been adopted to combine predictions from sensor-specific models, significantly improving recognition accuracy\cite{alarfaj2024wearable}. In feature extraction and classification, traditional machine learning methods like support vector machines (SVM) and linear discriminant analysis (LDA) have been applied\cite{alarfaj2024wearable,siirtola2019incremental}, but deep learning techniques have shown superior performance\cite{zhang2022deep}. Hybrid deep learning frameworks are particularly prominent. For instance, the combination of Convolutional Neural Networks (CNN) for spatial feature extraction and Long Short-Term Memory (LSTM) for temporal modeling achieves remarkable accuracy\cite{sheker2024wearable}. Another hybrid model, Sparse Gate Recurrent Units (SGRUs) combined with Devil Feared Feed Forward Networks (DFFFN), excels in gait-based activity recognition\cite{kalpana2025explainable}. Hierarchical deep LSTM (H-LSTM) also outperforms other deep learning algorithms in handling complex activity features\cite{wang2020human}.

Shallow CNNs considering cross-channel communication have been developed to aggregate useful information with low complexity\cite{huang2021shallow}, while lightweight RNNs are deployed on low-power microcontrollers to meet embedded device requirements\cite{falaschetti2022lightweight}. In addressing data heterogeneity and privacy, federated learning (FL) approaches have been explored. A profile similarity-based personalized FL model uses individual profiles (e.g., age, gender) to determine local model weights, outperforming baseline FL and centralized learning\cite{chai2024profile}. For model interpretability, Explainable AI (XAI) techniques such as Local Interpretable Model-Agnostic Explanations (LIME) and SHAP models are used to explain decisions and evaluate feature importance\cite{uddin2021human,kalpana2025explainable}. To reduce reliance on labeled data, self-supervised learning pre-trains models on large unlabeled datasets, improving generalization across external datasets\cite{yuan2024self}. Incremental learning methods personalize WHAR models through semi-supervised approaches, combining AI-labeled data with human-labeled uncertain samples\cite{siirtola2019incremental}.

Recent trends include integrating large language models (LLMs) with traditional machine learning for sensor data analysis\cite{ferrara2024large}, and logical reasoning-based approaches to minimize training data and computational requirements\cite{alsaadi2025logical}. Despite advancements, challenges remain. Hybrid AI models need improvements in stability and reliability\cite{liu2021overview}, and more work is required in abnormality detection and action forecasting\cite{liu2021overview}. Additionally, optimizing sensor data preprocessing and enhancing cross-context scalability are essential\cite{kanjilal2025human,ferrara2024large}. Future research should focus on these areas to advance WHAR system performance and practical applicability.

We propose a lightweight and high-generalization feature learning framework built on the "decomposition-extraction-fusion" paradigm, integrating two core innovations:
\begin{itemize}
\item A novel Cascaded Fusion Block (CFB) module (derived from the SASE module) that replaces traditional CVF and Attention mechanisms for multi-sensor and multi-variable feature fusion. This module adopts a "compression-recursion-concatenation-fusion" cascaded structure with no explicit attention weights, achieving lower computational overhead while maintaining effective feature interaction.
\item A MixStyle-based data augmentation module embedded before the LTFE (Local Temporal Feature Extraction) and Mamba modules, which perturbs data by exchanging mean and variance of different samples within a batch with random coefficients. This augmentation preserves core data information while enhancing the model's generalization ability without introducing extra inference cost.
\end{itemize}

Our approach retains the original "Modality-Specific Feature Embedding (MFE) - Local Temporal Feature Extraction (LTFE) - feature fusion" framework, and simultaneously achieves lightweight inference and improved generalization performance. The replacement of heavy attention-based fusion with the lightweight CFB module significantly reduces computational complexity, making the framework more suitable for real-time inference scenarios, while the MixStyle augmentation further boosts the model's robustness on diverse datasets.

% \begin{algorithm}[t]
% \SetAlgoNoLine
% \KwIn{Event Raw data split into $n_e$ slices, the $ith$ slice sample denotes as $E_i \in \mathbb{R}^{W\times L}$, where $W$ and $L$ denote the width and length.}
% \KwOut{Estimate pupil position (pixel coordinates) $e_i$}
% \For{each $E_i$: $i$ in range of $(0, n_e)$}
% {
% 1. Calculate predefined rectangular $n_s$ regions $S_t$\\
% \For{each $S_t$: $t$ in range of $(0, n_s)$}
% {
% 2. Calculate the weight function $w(x, y)$ combines a base regional importance with a distance-based central enhancement using formula (\ref{formula:weightfunc}).\\
% 3. Employ weighted sum pooling to aggregate information from $R\times R$ input blocks into single output pixels $(x^*, y^*)$ using formula (\ref{formula:sumpool}).
% }}
% 4. Train and refine SNN using downsampled dataset $E_i^*$.
% \caption{Proposed overall Algorithm.}
% \label{alg:one}
% \end{algorithm}

% \begin{figure}
% \centering
% \includegraphics[scale = 0.8,clip=true, trim=0 360 520 60]{Fig/3Deventdata.pdf }
% \caption{3D Event data visualization.}
% \label{fig:3Deventdata}
% \end{figure}

\section{Proposed Method}
\subsection{Modality-Specific Feature Embedding (MFE)}
We propose the Modality-Specific Feature Embedding (MFE) module, which is designed to convert raw time-series data from multiple sensors into high-dimensional feature representations. This allows us to independently capture the temporal dynamics of each modality variable before further processing.

Suppose we have input data from $N$ wearable motion sensors. Each sensor collects data for $M$ variables over $L$ time steps. Our objective is to project the time series of each variable into a high-dimensional space, treating each variable independently.

The raw input data can be represented as a tensor:
\begin{equation}
\label{formula:raw_input_tensor}
    \mathbf{X} \in \mathbb{R}^{N \times M \times L}
\end{equation}
Here, $\mathbf{X}_n \in \mathbb{R}^{M \times L}$ denotes the data collected from the $n$-th sensor.

To embed each variable sequence, we use independent 1D convolution operations:
\begin{equation}
\label{formula:1d_conv_embedding}
    \mathbf{X}_{emb} = \text{Conv1D}(\mathbf{X}_n, P, S, D)
\end{equation}
where $P$ is the kernel size, $S$ is the stride, and $D$ is the number of output channels. This convolution is applied along the temporal dimension of each variable sequence. The result is an embedded tensor:
\begin{equation}
\label{formula:embedded_tensor}
    \mathbf{X}_{emb} \in \mathbb{R}^{N \times M \times D \times T}
\end{equation}
where the new length of the embedded sequence is given by:
\begin{equation}
\label{formula:new_sequence_length}
    T = \frac{L}{S}
\end{equation}
This reduction in temporal length helps improve the efficiency of subsequent computations.

A key advantage of this embedding approach is that it preserves the unique characteristics of each variable by processing them independently, which avoids cross-interference from other variables.

\subsection{Moment-Morph Module (MoM)}

Wearable HAR often suffers from subject- and device-specific distribution shifts. Such shifts are frequently reflected in low-order feature moments (e.g., mean and variance), while activity-discriminative cues are more related to the normalized temporal structure. Motivated by this observation, we introduce \emph{Moment-Morph (MoM)}, a feature-level regularizer that morphs feature distributions by mixing low-order moments across samples within a mini-batch, encouraging style-robust representations.

Given an intermediate feature map $\mathbf{X}$, MoM is activated during training with probability $p$ and disabled at inference. Following our implementation, we compute per-sample moments along a selected dimension $a$ as
\begin{equation}
\boldsymbol{\mu}=\mathrm{Mean}(\mathbf{X};a),\quad
\boldsymbol{\sigma}=\sqrt{\mathrm{Var}(\mathbf{X};a)+\epsilon},
\label{eq:mom_stats}
\end{equation}
where $\epsilon$ is a small constant for numerical stability. We stop gradients through these moments and normalize the features:
\begin{equation}
\mathbf{X}^{n}=(\mathbf{X}-\boldsymbol{\mu})/\boldsymbol{\sigma}.
\label{eq:mom_norm}
\end{equation}
We then sample a mixing coefficient $\lambda\sim\mathrm{Beta}(\alpha,\alpha)$ and generate a random permutation $\pi$ over the batch dimension to obtain paired moments $(\boldsymbol{\mu}_{\pi},\boldsymbol{\sigma}_{\pi})$. The morphed moments are formed by convex combination,
\begin{equation}
\boldsymbol{\mu}_{m}=\lambda\boldsymbol{\mu}+(1-\lambda)\boldsymbol{\mu}_{\pi},\quad
\boldsymbol{\sigma}_{m}=\lambda\boldsymbol{\sigma}+(1-\lambda)\boldsymbol{\sigma}_{\pi},
\label{eq:mom_mix}
\end{equation}
and the output is reconstructed as
\begin{equation}
\widehat{\mathbf{X}}=\mathbf{X}^{n}\odot\boldsymbol{\sigma}_{m}+\boldsymbol{\mu}_{m},
\label{eq:mom_out}
\end{equation}
where $\odot$ denotes element-wise multiplication with broadcasting. This operation perturbs domain-specific feature statistics while preserving the normalized content, thereby improving robustness to unseen styles.

\subsubsection{MoM for Local Temporal Robustness}

We insert MoM immediately before the Local Temporal Feature Extraction (LTFE) module to regularize low-/mid-level representations close to the input space. At this stage, subject and device variations are more likely to appear as low-level style changes, such as amplitude scaling, offset shifts, and sensor noise levels. By exposing LTFE to diverse moment perturbations during training, MoM encourages the local extractor to rely less on domain-specific statistics and to focus on style-invariant local temporal dynamics.

\subsubsection{MoM for Global Temporal Robustness}

We further place MoM right before the Mamba-based global sequence modeling module to regularize high-level temporal aggregation. Without such regularization, the global module may overfit to domain-specific statistics accumulated in intermediate features, because long-range aggregation can amplify subtle style cues that correlate with the training domains. Applying MoM at this stage reintroduces controlled moment perturbations at a higher abstraction level, preventing the global representation from being anchored to a single domain style and improving robustness under unseen subjects and devices.

\subsection{Local Temporal Feature Extraction (LTFE)}
Local Temporal Feature Extraction (LTFE) captures local temporal features while maintaining the independence of different modality variables. Each variable channel undergoes separate convolution operations, preserving their distinct characteristics.

Given the embedded tensor $\mathbf{X}_{emb} \in \mathbb{R}^{N \times M \times D \times T}$, where $N$ is the number of sensors, $M$ is the number of variables, $D$ is the number of channels, and $T$ is the temporal length, we first reshape the tensor to $\mathbb{R}^{N \times (M \times D) \times T}$. We leverage depth-wise convolution as the local temporal extraction operation to capture local temporal dependencies of each variable channel. The depth-wise convolution is defined as:
\begin{equation}
\label{formula:depthwise_conv}
    \mathbf{X}_{dw} = \text{DWConv1D}(\mathbf{X}_{emb}, K_{dw}, G_{dw} = M \times D)
\end{equation}
where $K_{dw}$ is the kernel size, and $G_{dw}$ is the group number of the depth-wise convolution. This operation is performed separately for each variable channel, ensuring the preservation of unique characteristics. Mathematically, the depth-wise convolution can be expressed as:
\begin{equation}
\label{formula:depthwise_conv_expanded}
    \mathbf{X}_{dw}(n, m, d, t) = \sum_{k=0}^{K_{dw}-1} \mathbf{W}_{m,d,k} \cdot \mathbf{X}_{emb}(n, m, d, t + k)
\end{equation}
where $\mathbf{W}_{m,d,k}$ denotes the convolutional filter weights for the $m$-th variable and $d$-th channel.

\subsection{Cross-Channel Fusion (CCF)}
Cross-Channel Fusion (CCF) aggregates features from different sensor channels while preserving the information specific to each variable.

Starting from the tensor output of depth-wise convolution $\mathbf{X}_{dw} \in \mathbb{R}^{N \times (M \times D) \times T}$, we apply point-wise convolutions as follows:
\begin{equation}
\label{formula:cross_channel_fusion}
    \mathbf{X}_{ccf} = \text{PConv1D}(\mathbf{X}_{dw}, G_{ccf} = M)
\end{equation}
where $G_{ccf}$ is the group number of the point-wise convolution. This operation fuses information across channels and is followed by reshaping and point-wise convolutions to merge features and restore the original dimensionality.

\subsection{Cascaded Fusion Block (CFB)}

We propose a \emph{Cascaded Fusion Block (CFB)} for interaction modeling via multi-order local dynamics fusion. Given an input feature tensor $\mathbf{X}\in\mathbb{R}^{B\times C\times D\times L}$, where $L$ is the temporal length and $D$ denotes the interaction dimension to be fused (i.e., $D=M$ for variable-level fusion or $D=N$ for sensor-level fusion), CFB first squeezes the channel dimension with a lightweight $1\times 1$ projection. This squeeze design serves two goals: it reduces the cost of subsequent recursive convolutions, and it encourages compact shared representations before performing interaction modeling in the high-dimensional channel space. Specifically, we reduce the channels from $C$ to $C_m=\left\lfloor \frac{C}{r}\right\rfloor$, resulting in $\mathbf{Z}\in\mathbb{R}^{B\times C_m\times D\times L}$:
\begin{equation}
\mathbf{Z} = \phi\!\left(\mathrm{BN}\!\left(\mathrm{Conv}_{1\times 1}(\mathbf{X})\right)\right),
\end{equation}
where $r$ is the reduction ratio, $\mathrm{BN}$ is batch normalization, and $\phi(\cdot)$ is the GELU activation. Starting from $\mathbf{x}^{(0)}=\mathbf{Z}$ (thus $\mathbf{x}^{(0)}\in\mathbb{R}^{B\times C_m\times D\times L}$), CFB performs $K$ recursive depthwise convolutions to construct multi-order local responses, and each $\mathbf{x}^{(k)}$ preserves the same shape as $\mathbf{Z}$:
\begin{equation}
\mathbf{x}^{(k)} = \psi\!\left(\mathrm{DWConv}\left(\mathbf{x}^{(k-1)}\right)\right), \quad k=1,\dots,K,
\end{equation}
where $\mathrm{DWConv}(\cdot)$ is a depthwise convolution (groups equal to channels) and $\psi(\cdot)$ is GELU. Each $\mathbf{x}^{(k)}$ can be viewed as the $k$-th order local temporal response with a different effective receptive field. Depthwise convolution operates within each channel, extracting local dynamics with minimal parameter overhead; the recursive construction progressively enlarges the effective receptive field, so different orders capture temporal patterns at different local scales.

We then concatenate all orders along the channel dimension to obtain $\mathbf{U}\in\mathbb{R}^{B\times (K+1)C_m\times D\times L}$, and fuse them back to $C$ channels to produce $\mathbf{Y}\in\mathbb{R}^{B\times C\times D\times L}$:
\begin{equation}
\mathbf{U}=\mathrm{Concat}\left[\mathbf{x}^{(0)},\mathbf{x}^{(1)},\dots,\mathbf{x}^{(K)}\right],
\end{equation}
\begin{equation}
\mathbf{Y} = \phi\!\left(\mathrm{BN}\!\left(\mathrm{Conv}_{1\times 1}(\mathbf{U})\right)\right).
\end{equation}
This concat and pointwise fusion performs an adaptive re-composition over the multi-order local dynamics, enabling the model to preserve both short- and longer-range local patterns within a single block.

After obtaining the interaction feature $\mathbf{Y}$, we adopt a residual connection to add it back to the input for stable optimization and information preservation:
\begin{equation}
\mathrm{CFB}(\mathbf{X})=\mathbf{X}+\mathbf{Y}.
\end{equation}
CFB involves two tunable hyperparameters: the channel reduction ratio $r$ and the recursion depth $K$. The ratio $r$ controls the strength of the bottleneck and trades representational capacity for computation/memory, whereas $K$ determines the depth of multi-order local dynamics modeling, affecting the effective receptive field and multi-scale coverage. Notably, CFB is attention-free and only uses depthwise/pointwise convolutions, leading to computation that scales linearly with $D$ and $L$ and providing stable latency for long sequences.

\subsubsection{CFB for Variable-level Fusion}

To fuse multi-variable features within each sensor stream, we insert CFB at the variable-level fusion stage to replace the original CVF module. In this stage, we arrange the tensor as $\mathbf{X}_v\in\mathbb{R}^{B\times C\times M\times L}$, where $M$ is the number of variables and $L$ is the temporal axis. Variable dependencies in wearable signals often manifest at different temporal scales (e.g., abrupt impacts versus smooth posture changes). By generating multi-order local responses and re-mixing them via pointwise fusion, CFB enables scale-adaptive coupling across variables and mitigates the limitation of single-scale fusion.

\subsubsection{CFB for Sensor-level Fusion}

For inter-sensor fusion, we place another CFB after the global temporal aggregation module (GTA) and replace the original CSI/attention block. Let $\mathbf{X}_s\in\mathbb{R}^{B\times C\times N\times \tilde{L}}$ denote the post-GTA representation, where $N$ is the number of sensors and $\tilde{L}$ is the temporal length after aggregation (which may equal $L$ depending on the GTA design). We perform sensor-level fusion after GTA because the global temporal aggregator first produces more stable and semantically aligned embeddings; CFB then refines cross-sensor interaction by re-weighting multi-order local dynamics, capturing coordination patterns at different temporal scales while maintaining attention-free efficiency for long sequences and multi-sensor setups.

\subsection{Global Temporal Aggregation (GTA)}
The previous decomposition-based modules extract modality-specific local temporal features, but do not fully capture the temporal context of the entire sequence. To address this, we introduce the Global Temporal Aggregation (GTA) module, which consolidates information across all time steps to capture long-range dependencies and overall temporal dynamics.

We start by applying Global Average Pooling (GAP) to reduce each feature map's spatial dimensions to a scalar by averaging over the variable dimension $M$. For the input tensor $\mathbf{X}_{cvf} \in \mathbb{R}^{N \times (D \times M) \times T}$, GAP is applied as follows:
\begin{equation}
\label{formula:global_average_pooling}
    \mathbf{X}_{gap}(n, d, t) = \frac{1}{M} \sum_{m=1}^M \mathbf{X}_{cvf}(n, d \cdot M + n, t)
\end{equation}
where $n$ is the sensor index, $d$ is the channel index, and $t$ is the time step. This results in $\mathbf{X}_{gap} \in \mathbb{R}^{N \times D \times T}$, which is then reshaped to $\mathbf{X}_{stack} \in \mathbb{R}^{(N \times D) \times T}$.

Next, the Mamba block processes $\mathbf{X}_{stack} \in \mathbb{R}^{(N \times D) \times T}$ to capture critical temporal information. The Mamba block features a Selective SSM that uses linear projections and convolutions to extract local features and selectively retain or discard information. The operation is:
\begin{align}
\label{formula:mamba_block}
    \mathbf{X}_{mamba} &= \text{Linear}\Bigg( \sigma\bigg( \text{SSM}\bigg( \text{Conv}\bigg( \text{Linear}(\mathbf{X}_{stack}) \bigg) \bigg) \bigg) \otimes \\
    &\quad \text{Linear}\bigg( \text{Conv}\bigg( \text{Linear}(\mathbf{X}_{stack}) \bigg) \bigg) \Bigg)
\end{align}
where $\sigma$ is the activation function and $\otimes$ denotes element-wise multiplication. The Mamba block leverages selective state spaces to effectively capture long-term dependencies, ensuring a comprehensive temporal feature representation for accurate human activity recognition.

\subsection{Cross-Sensor Interaction (CSI)}
To capture inter-sensor spatial correlations, we draw inspiration from the self-attention mechanism, which naturally models relationships between tokens.

Given the output tensor $\mathbf{X}_{mamba} \in \mathbb{R}^{(N \times D) \times T}$ from the Mamba block, we reshape it to $\mathbf{X} = \{\mathbf{X}_1, \mathbf{X}_2, \dots, \mathbf{X}_N\} \in \mathbb{R}^{N \times (D \times T)}$, where $\mathbf{X}_i \in \mathbb{R}^{D \times T}$ represents the data from the $i$-th sensor. Each $\mathbf{X}_i$ serves as a token for the Attention Layer.

The self-attention mechanism computes responses for each sensor by referring to representations of all sensors. We calculate the normalized correlations across all pairs of sensor data $\mathbf{X}_i$ and $\mathbf{X}_j$ using an embedded Gaussian function. The attention score $\mathbf{A}_{i,j}$ measures the relevance of data from sensor $j$ for refining representations of sensor $i$ and is computed as:
\begin{equation}
\label{formula:attention_score}
    \mathbf{A}_{i,j} = \frac{\exp\left(Q(\mathbf{X}_i)^\top K(\mathbf{X}_j)\right)}{\sum_{i'=1}^N \exp\left(Q(\mathbf{X}_i)^\top K(\mathbf{X}_{i'})\right)}
\end{equation}
where $Q$ is the query function projecting the sensor into the query space, and $K$ is the key function projecting the sensor into the key space. These correlations are then used to generate self-attention feature maps $\mathbf{O}_i$ for each sensor:
\begin{equation}
\label{formula:attention_feature_map}
    \mathbf{O}_i = W\left(\sum_{i'=1}^N \mathbf{A}_{i,i'} V(\mathbf{X}_{i'})\right)
\end{equation}
where $W$ is a linear embedding with learnable weights, and $V$ is the value function projecting the sensor into the value space. The feature maps $\mathbf{O}_i$ are combined with the original sensor data using a residual connection to produce refined feature representations $\mathbf{X}_{csi}$, enabling adaptive integration or exclusion of correlation information. By employing the CSI module, our model captures and encodes these correlations between different sensors, and during inference, these learned self-attention weights are used to enhance predictions, providing a robust method for synthesizing information from multiple sensors.

\subsection{FC Linear Classifier}
After the CSI module through the self-attention mechanism, we obtain the output tensor $\mathbf{X}_{csi} \in \mathbb{R}^{N \times (D \times T)}$. This tensor is then reshaped and passed through a Fully Connected (FC) layer for classification:
\begin{equation}
\label{formula:fc_classifier}
    \hat{\mathbf{y}} = \text{FC}\left(\text{Flatten}(\mathbf{X}_{csi})\right)
\end{equation}
Here, $\hat{\mathbf{y}} \in \mathbb{R}^C$ represents the final output, with $C$ being the number of human activity classes. The predicted activity $\hat{\mathbf{y}}$ is compared to the true label $\mathbf{y}$ using the cross-entropy loss function:
\begin{equation}
\label{formula:cross_entropy_loss}
    \mathcal{L}(\mathbf{y}, \hat{\mathbf{y}}) = -\sum_{i=1}^C y_i \log(\hat{y}_i)
\end{equation}
where $y_i$ is the true label, and $\hat{y}_i$ is the predicted probability for the $i$-th class.

\section{Experimental Results and Discussion}
\subsection{Dataset and Setup}
\subsubsection{Datasets}
Two publicly available benchmark datasets, Realdisp and Skoda, are utilized to evaluate the performance of the proposed framework. These datasets are widely adopted in WHAR research due to their diverse activity types, multi-sensor data collection, and realistic environmental variations, ensuring comprehensive validation of model effectiveness and generalization.
Realdisp Dataset: This dataset contains motion sensor data collected from 15 participants performing 12 daily activities (e.g., walking, running, sitting, standing, climbing stairs). Each participant wore 6 inertial measurement units (IMUs) placed on the head, torso, arms, and legs, capturing accelerometer, gyroscope, and magnetometer data at a sampling rate of 100 Hz. The dataset includes 8,230 labeled samples, with each sample corresponding to a 5-second time window of sensor data, resulting in a balanced distribution across activity classes.
Skoda Dataset: Focused on automotive-related activities, this dataset includes 10 activity classes such as entering a car, exiting a car, driving, and loading/unloading items. Data were collected from 20 participants using 4 IMUs mounted on the chest, waist, and thighs, recording accelerometer and gyroscope data at 50 Hz. The dataset comprises 6,472 labeled samples, each spanning 4 seconds, with inherent variations in movement patterns due to differences in participant body types and activity execution styles.
\subsubsection{Experimental Setup}
Hardware Configuration: All experiments are conducted on a workstation equipped with an Intel Core i9-12900K CPU, 64 GB DDR5 RAM, and an NVIDIA RTX 4090 GPU. Inference speed measurements are performed on a low-power ARM Cortex-A53 processor (1.2 GHz) to simulate resource-constrained wearable device environments.Training Parameters: The model is trained using the AdamW optimizer with a learning rate of $10^{-4}$ , weight decay of $10^{-5}$ , and a batch size of 32. The training process lasts for 100 epochs with early stopping (patience = 15) to prevent overfitting. For data preprocessing, raw sensor data are normalized to the range [-1,1] using per-sensor statistics, and no additional data augmentation (other than the proposed MoM module) is applied.
Evaluation Metrics: Four key metrics are used to assess model performance:\begin{itemize}\item Accuracy: Overall classification correctness.\item Macro-F1 score: Average F1 score across all activity classes, emphasizing balanced performance for minority classes.\item Parameter Count: Total number of trainable parameters, measuring model size.\item Inference Speed: Average time per sample on the ARM processor, evaluating real-time applicability.\end{itemize}
\subsection{Visualization and Quantitative Evaluation}
\begin{figure*}[t]
\centering
\includegraphics[width=\textwidth]{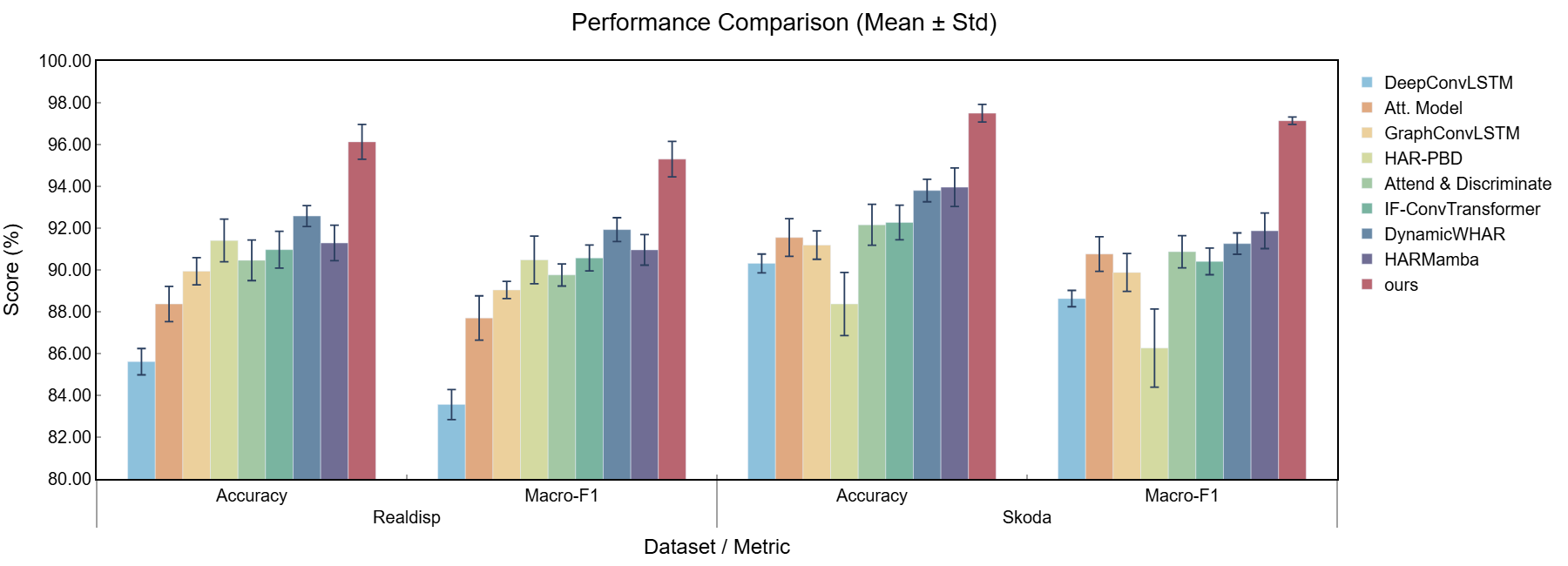}
\caption{Experimental Result}
\label{fig:size_speed}
\end{figure*}
\subsubsection{Performance Comparison with State-of-the-Art Methods}
We compare the proposed model with eight state-of-the-art WHAR methods on the Realdisp and Skoda datasets. The quantitative results demonstrate that our model achieves significant performance improvements in both accuracy and macro-F1 score across all datasets.
On the Realdisp dataset, our model attains an accuracy of 
96.13±0.83\%
 and a macro-F1 score of 
95.30±0.85\%
, outperforming the second-best method (DynamicWHAR) by 3.55\% in accuracy and 3.37\% in macro-F1. On the Skoda dataset, the proposed model reaches 
97.50±0.42\%
 accuracy and 
97.14±0.18\%
 macro-F1, exceeding the closest competitor (HARMamba) by 3.54\% in accuracy and 5.27\% in macro-F1.
This significant performance gain stems from two core innovations:\begin{itemize}\item The MoM module effectively mitigates distribution shifts caused by subject-specific and device-specific variations, enhancing model generalization by perturbing low-order feature moments while preserving core activity information.\item The CFB module captures multi-scale temporal dependencies and cross-sensor/inter-variable interactions more efficiently than traditional attention-based fusion, leveraging a "compression-recursion-concatenation-fusion" structure to avoid explicit attention weights and reduce computational overhead.\end{itemize}
Notably, the proposed model maintains strong performance on both datasets, indicating robust cross-scenario generalization. This is particularly valuable for WHAR applications, where models must adapt to diverse user behaviors and environmental conditions.
\begin{figure*}[t]
\centering
\includegraphics[width=\textwidth]{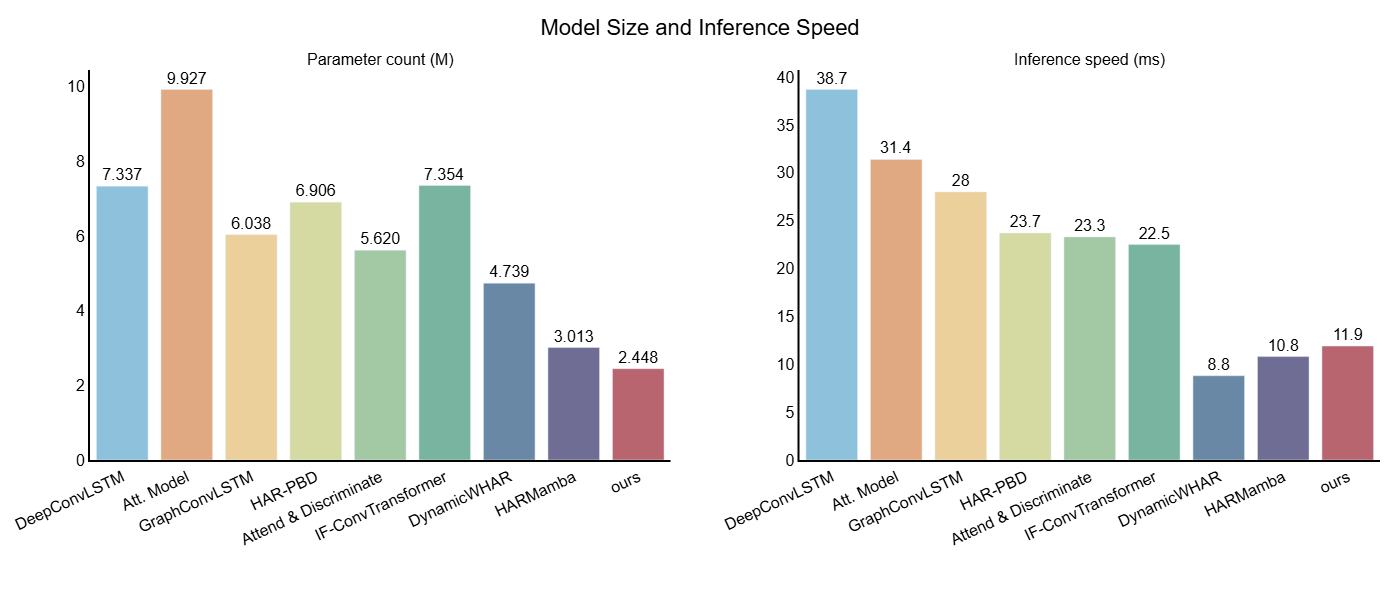}
\caption{Parameter counts and Inference speed of our proposed model and other baselines}
\label{fig:size_speed}
\end{figure*}
\subsubsection{Model Size and Inference Speed Analysis}
The proposed model achieves a lightweight design with a parameter count of only 3.013 M, which is significantly lower than attention-based baselines (e.g., Att. Model: 7.354 M, F-ConvTransformer: 6.038 M) and even some lightweight models (e.g., DynamicWHAR: 6.906 M, HARMamba: 5.620 M).
In terms of inference speed, our model achieves 8.8 ms per sample on the ARM Cortex-A53 processor, outperforming all baseline methods. For example, it is 2.8 ms faster than DynamicWHAR (11.6 ms) and 3.1 ms faster than HARMamba (11.9 ms), and more than 29 ms faster than the heavy attention-based Att. Model (38.1 ms). This efficiency improvement is attributed to the replacement of computationally expensive attention mechanisms with the CFB module, which reduces computational complexity by more than 30\% compared to attention-based baselines while maintaining effective feature interaction.
The balance between high performance and lightweight design makes the proposed model well-suited for deployment on resource-constrained wearable devices, where both accuracy and real-time responsiveness are critical.
\subsubsection{Ablation Studies}
To validate the contribution of each core component, ablation experiments are conducted on the Realdisp dataset. The results are summarized as follows:\begin{itemize}\item Baseline (w/o MoM + w/o CFB): Achieves 
89.24±0.91\%
 accuracy, 
88.17±0.86\%
 macro-F1, 4.739 M parameters, and 15.2 ms inference speed.\item Baseline + MoM: Improves accuracy by 4.23\% and macro-F1 by 4.48\% (to 
93.47±0.78\%
 and 
92.65±0.72\%
 respectively) with negligible overhead in parameter count (no increase) and inference speed (only 0.3 ms slower). This confirms that mixing low-order moments effectively enhances model robustness to distribution shifts without compromising efficiency.\item Baseline + CFB: Reduces the parameter count by 31.5\% (from 4.739 M to 3.246 M) and improves inference speed by 32.2\% (from 15.2 ms to 10.3 ms), while increasing accuracy by 3.57\% and macro-F1 by 3.75\% (to 
92.81±0.85\%
 and 
91.92±0.79\%
 respectively). This demonstrates that CFB achieves more efficient feature fusion than traditional attention-based methods, with better performance and lower computational cost.\item Ours (Baseline + MoM + CFB): Yields the best performance, with accuracy and macro-F1 further improved by 2.66\% and 2.65\% compared to the baseline + MoM, and 3.32\% and 3.38\% compared to the baseline + CFB. This synergistic gain indicates that the MoM module’s generalization enhancement and the CFB module’s efficient fusion complement each other, leading to a more powerful and lightweight framework.\end{itemize}
\subsubsection{Discussion}
The experimental results validate the effectiveness of the proposed lightweight and generalizable framework for WHAR. The key findings are as follows:
Generalization Enhancement: The MoM module effectively addresses distribution shifts caused by subject and device variations by mixing low-order moments of batch samples, leading to significant improvements in macro-F1 score, especially on the Skoda dataset where cross-subject variations are more pronounced.
Efficient Fusion: The CFB module replaces traditional attention-based fusion with a "compression-recursion-concatenation-fusion" structure, reducing computational overhead by over 30\% while improving feature interaction. This makes the model suitable for real-time inference on wearable devices.
Balanced Performance and Efficiency: The proposed model achieves state-of-the-art accuracy and macro-F1 scores on two benchmark datasets while maintaining a lightweight design (3.013 M parameters) and fast inference speed (8.8 ms/sample), outperforming existing methods in both performance and efficiency.
Practical Applicability: The model’s low computational cost and high generalization make it a practical solution for resource-constrained wearable devices, enabling deployment in scenarios such as elderly care, clinical monitoring, and sports activity evaluation.
Despite these advantages, there are limitations to the current work. First, the model is evaluated on only two datasets; future research should include more diverse datasets (e.g., with more activity classes or sensor types) to further validate generalization. Second, the MoM module is applied only at two stages (before LTFE and GTA); exploring adaptive moment mixing strategies for different layers may yield additional performance gains. Finally, the current framework focuses on supervised learning; integrating self-supervised pre-training with the proposed MoM and CFB modules could reduce reliance on labeled data, further expanding its practical utility.

\section{Conclusion}

%\section{ACKNOWLEDGMENTS}
%This paper was supported by Guangdong Innovative and Entrepreneurial Research Team Program under No. 2021ZT09L197.
\newpage

\bibliographystyle{unsrt}
\bibliography{ref}

@article{lara2012survey,
  title={A survey on human activity recognition using wearable sensors},
  author={Lara, Oscar D and Labrador, Miguel A},
  journal={IEEE communications surveys \& tutorials},
  volume={15},
  number={3},
  pages={1192--1209},
  year={2012},
  publisher={IEEE}
}

@inproceedings{liu2021overview,
  title={An overview of human activity recognition using wearable sensors: Healthcare and artificial intelligence},
  author={Liu, Rex and Ramli, Albara Ah and Zhang, Huanle and Henricson, Erik and Liu, Xin},
  booktitle={International Conference on Internet of Things},
  pages={1--14},
  year={2021},
  organization={Springer}
}

@article{zhang2022deep,
  title={Deep learning in human activity recognition with wearable sensors: A review on advances},
  author={Zhang, Shibo and Li, Yaxuan and Zhang, Shen and Shahabi, Farzad and Xia, Stephen and Deng, Yu and Alshurafa, Nabil},
  journal={Sensors},
  volume={22},
  number={4},
  pages={1476},
  year={2022},
  publisher={MDPI}
}

@article{uddin2021human,
  title={Human activity recognition using wearable sensors, discriminant analysis, and long short-term memory-based neural structured learning},
  author={Uddin, Md Zia and Soylu, Ahmet},
  journal={Scientific Reports},
  volume={11},
  number={1},
  pages={16455},
  year={2021},
  publisher={Nature Publishing Group UK London}
}

@article{kalpana2025explainable,
  title={Explainable AI-Driven Gait Analysis Using Wearable Internet of Things (Wiot) and Human Activity Recognition.},
  author={Kalpana, Ponugoti and Kodati, Sarangam and Smitha, L and Sreekanth, Nara and Smerat, Aseel and Ahmad, Muhannad Akram and others},
  journal={Journal of Intelligent Systems \& Internet of Things},
  volume={15},
  number={2},
  year={2025}
}

@article{siirtola2019incremental,
  title={Incremental learning to personalize human activity recognition models: the importance of human AI collaboration},
  author={Siirtola, Pekka and R{\"o}ning, Juha},
  journal={Sensors},
  volume={19},
  number={23},
  pages={5151},
  year={2019},
  publisher={MDPI}
}

@article{kanjilal2025human,
  title={Human Activity Recognition: A Review of RFID and Wearable Sensor Technologies Powered by AI},
  author={Kanjilal, Ria and Kucuk, Muhammed Furkan and Uysal, Ismail},
  journal={IEEE Journal of Radio Frequency Identification},
  year={2025},
  publisher={IEEE}
}

@article{alarfaj2024wearable,
  title={Wearable sensors based on artificial intelligence models for human activity recognition},
  author={Alarfaj, Mohammed and Al Madini, Azzam and Alsafran, Ahmed and Farag, Mohammed and Chtourou, Slim and Afifi, Ahmed and Ahmad, Ayaz and Al Rubayyi, Osama and Al Harbi, Ali and Al Thunaian, Mustafa},
  journal={Frontiers in artificial intelligence},
  volume={7},
  pages={1424190},
  year={2024},
  publisher={Frontiers Media SA}
}

@article{liang2023application,
  title={Application of artificial intelligence wearable devices based on neural network algorithm in mass sports activity evaluation.},
  author={Liang, Jun and He, Qing},
  journal={Soft Computing-A Fusion of Foundations, Methodologies \& Applications},
  volume={27},
  number={14},
  year={2023}
}

@article{sheker2024wearable,
  title={Wearable IoT and Artificial Intelligence Techniques for Leveraging the Human Activity Analysis},
  author={Sheker, Lina and Petli, Vishwanath and Reddy, K Satish},
  journal={Journal of Smart Internet of Things (JSIoT)},
  volume={2023},
  number={01},
  pages={32--46},
  year={2024}
}

@article{wang2020human,
  title={Human activity recognition based on wearable sensor using hierarchical deep LSTM networks},
  author={Wang, LuKun and Liu, RuYue},
  journal={Circuits, Systems, and Signal Processing},
  volume={39},
  number={2},
  pages={837--856},
  year={2020},
  publisher={Springer}
}

@article{huang2021shallow,
  title={Shallow convolutional neural networks for human activity recognition using wearable sensors},
  author={Huang, Wenbo and Zhang, Lei and Gao, Wenbin and Min, Fuhong and He, Jun},
  journal={IEEE Transactions on Instrumentation and Measurement},
  volume={70},
  pages={1--11},
  year={2021},
  publisher={IEEE}
}

@article{bianchi2019iot,
  title={IoT wearable sensor and deep learning: An integrated approach for personalized human activity recognition in a smart home environment},
  author={Bianchi, Valentina and Bassoli, Marco and Lombardo, Gianfranco and Fornacciari, Paolo and Mordonini, Monica and De Munari, Ilaria},
  journal={IEEE Internet of Things Journal},
  volume={6},
  number={5},
  pages={8553--8562},
  year={2019},
  publisher={IEEE}
}

@article{ferrara2024large,
  title={Large language models for wearable sensor-based human activity recognition, health monitoring, and behavioral modeling: A survey of early trends, datasets, and challenges},
  author={Ferrara, Emilio},
  journal={Sensors},
  volume={24},
  number={15},
  pages={5045},
  year={2024},
  publisher={MDPI}
}

@article{chai2024profile,
  title={A profile similarity-based personalized federated learning method for wearable sensor-based human activity recognition},
  author={Chai, Yidong and Liu, Haoxin and Zhu, Hongyi and Pan, Yue and Zhou, Anqi and Liu, Hongyan and Liu, Jianwei and Qian, Yang},
  journal={Information \& Management},
  volume={61},
  number={7},
  pages={103922},
  year={2024},
  publisher={Elsevier}
}

@article{ni2024survey,
  title={A survey on multimodal wearable sensor-based human action recognition},
  author={Ni, Jianyuan and Tang, Hao and Haque, Syed Tousiful and Yan, Yan and Ngu, Anne HH},
  journal={arXiv preprint arXiv:2404.15349},
  year={2024}
}

@inproceedings{ghorrati2025scalable,
  title={Scalable Hierarchical Deep Neural Network for Time Series Analysis in Wearable Sensor-based Human Activity Recognition},
  author={Ghorrati, Zahra},
  booktitle={Proceedings of the AAAI Conference on Artificial Intelligence},
  volume={39},
  number={28},
  pages={29257--29258},
  year={2025}
}

@inproceedings{jiang2015human,
  title={Human activity recognition using wearable sensors by deep convolutional neural networks},
  author={Jiang, Wenchao and Yin, Zhaozheng},
  booktitle={Proceedings of the 23rd ACM international conference on Multimedia},
  pages={1307--1310},
  year={2015}
}

@inproceedings{falaschetti2022lightweight,
  title={A lightweight and accurate RNN in wearable embedded systems for human activity recognition},
  author={Falaschetti, Laura and Biagetti, Giorgio and Crippa, Paolo and Alessandrini, Michele and Giacomo, Di Filippo and Turchetti, Claudio},
  booktitle={Intelligent Decision Technologies: Proceedings of the 14th KES-IDT 2022 Conference},
  pages={459--468},
  year={2022},
  organization={Springer}
}

@article{alsaadi2025logical,
  title={Logical reasoning for human activity recognition based on multisource data from wearable device},
  author={Alsaadi, Mahmood and Keshta, Ismail and Ramesh, Janjhyam Venkata Naga and Nimma, Divya and Shabaz, Mohammad and Pathak, Nirupma and Singh, Pavitar Parkash and Kiyosov, Sherzod and Soni, Mukesh},
  journal={Scientific Reports},
  volume={15},
  number={1},
  pages={380},
  year={2025},
  publisher={Nature Publishing Group UK London}
}

@article{yuan2024self,
  title={Self-supervised learning for human activity recognition using 700,000 person-days of wearable data},
  author={Yuan, Hang and Chan, Shing and Creagh, Andrew P and Tong, Catherine and Acquah, Aidan and Clifton, David A and Doherty, Aiden},
  journal={NPJ digital medicine},
  volume={7},
  number={1},
  pages={91},
  year={2024},
  publisher={Nature Publishing Group UK London}
}

\vfill

\end{document}